\documentclass[prd,aps,twocolumn,nofootinbib]{revtex4}
\usepackage{graphicx}
\usepackage[colorlinks]{hyperref}
\usepackage{xspace}
\usepackage{mathtools}
\usepackage{amsmath}

\def\xmm{\textit{XMM-Newton}\xspace}
\def\extp{\textit{eXTP}\xspace}

\begin{document}

\title{eXTP Perspectives for the $\nu$MSM Sterile Neutrino Dark Matter Model}
\author{Denys Malyshev$^1$, Charles Thorpe-Morgan$^1$, Andrea Santangelo$^1$, Josef Jochum$^1$, Shuang-Nan Zhang$^{2,3,4}$}
\address{$^{1}$ Institut f{\"u}r Astronomie und Astrophysik T{\"u}bingen, Universit{\"a}t T{\"u}bingen, Sand 1, D-72076 T{\"u}bingen, Germany\\
$^2$ Key Laboratory of Particle Astrophysics, Institute of High Energy Physics, Chinese Academy of Sciences, Beijing 100049, China\\
$^3$ University of Chinese Academy of Sciences, Beijing 100049, China\\
$^4$ National Astronomical Observatories of China, Chinese Academy of Sciences, Beijing 100012, China}
\begin{abstract}
We discuss the potential of the \extp X-ray telescope, in particular its Spectroscopic Focusing Array (SFA), Large Area Detector (LAD) and Wide Field Monitor (WFM) for the detection of a signal from keV-scale decaying dark matter. We show that the sensitivity of the \extp is sufficient to improve existing constraints on the mixing angle of the neutrino Minimal extension of the Standard Model ($\nu$MSM) by a factor of 5-10 within the dark matter mass range 2--50~keV, assuming 1\% level of systematic uncertainty. We assert that the \extp will be able to probe previously inaccessible range of $\nu$MSM parameters and serve as a precursor for the Athena mission in decaying dark matter searches. 
\end{abstract}

\maketitle

\section{Introduction}
\label{sec:intro}
Astrophysical sources offer attractive laboratories for testing and constraining the properties of dark matter(DM) through indirect detection of its annihilation or decay products (e.g. photons, neutrinos, charged particles). With the lack of any firm detection so far, the search remains ongoing and will be aided by the next generation of satellites. These future missions will allow access to previously unavailable sensitivities in search of DM, enabling better constraints of DM properties or, finally, measurements of its parameters. 

The lowest mass range for fermionic dark matter is known to be located in the keV band~\cite{tremainegunn,dodelson,tg1,savchenko19}. Several extensions of the Standard Model (SM) of particle physics incorporate dark-matter candidate particles which can produce radiative signatures in this band, including gravitinos~\cite{hamaguchi14}, pseudo-Nambu–Goldstone bosons~\cite{nakayama14}, axions and axion-like particles~\cite{arias12,henning15}. In what is below, we focus on one of the most well explored of such theories -- the minimal sterile neutrino extension of the SM ($\nu$MSM)~\cite{numsm11,numsm,numsm1,review_old,sterile_neutrino_review19}.

A sterile neutrino of mass $m_{DM}$ can decay producing a Standard Model neutrino and a monochromatic keV photon with an energy of $E=m_{DM}/2$~\cite{kev01,pal,sterile_neutrino_wp,sterile_neutrino_review19}. This decay signal can appear as a narrow line-like feature in X-ray spectra of astrophysical DM-dominated objects~\cite{kev01}, e.g. clusters of galaxies or dwarf spheroidal galaxies (dSphs). The strength of the signal is determined by the active-sterile neutrino mixing angle $\theta$.

The parameters of the $\nu$MSM model (the mass of sterile neutrino $m_{DM}$ and mixing angle $\theta$) are constrained from below and above and only a narrow window of the parameter space remains unexcluded so far, see e.g. Fig.~\ref{fig:dm_limits} and \cite{sterile_neutrino_review19} for a recent review.

The lower bound on the mass of fermionic dark matter particles $m_{DM}\gtrsim 1$~keV arises from limits imposed by the uncertainty relation. Specifically, the phase space density of the DM particles in the halos of dwarf spheroidal galaxies cannot exceed the fundamental limits imposed by the uncertainty relation and the initial phase space density at the moment of production of the DM in the Early Universe~\cite{tremainegunn,tg1,gorbunov08,savchenko19}.

High values of mixing angle $\theta$ are forbidden because the abundance of sterile neutrinos produced in the Early Universe with such mixing angles would exceed the observed DM density in the present day~(see e.g. \cite{dodelson,numsm1} and~\cite{sterile_neutrino_review19} for a recent review). Additional upper limits originate from non-detection of the described line-like feature in multiple DM-dominated objects with the current generation of instruments~\cite{sterile_neutrino_review19}.

The lower bound on the mixing angle indicates the region where the lepton asymmetries required for resonantly enhanced thermal sterile neutrino production to work are ruled out. Mixing angles lower than this bound would result in the abundances of light elements produced during Big Bang Nucleosynthesis to be in disagreement with the current measured values~\cite{shi99,serpico05,shaposhnikov08,laine08,canetti13}. Note however that these limits can be substantially relaxed in other production models, including e.g., Higgs decay~\cite{kusenko06}. 

In the following, we study the capabilities of the forthcoming \extp mission to probe the  remaining ``island'' of the allowed parameter range of the $\nu$MSM model, which is unexplored by the current generation X-ray instruments. Namely, we propose deep observations of a  DM-dominated object (dwarf spheroidal galaxy) and blank sky regions aiming either to detect the line from decaying dark matter, or to constrain $(m_{DM}, \theta)$ sterile neutrino parameters.

\begin{figure*}
\includegraphics[width=0.45\linewidth]{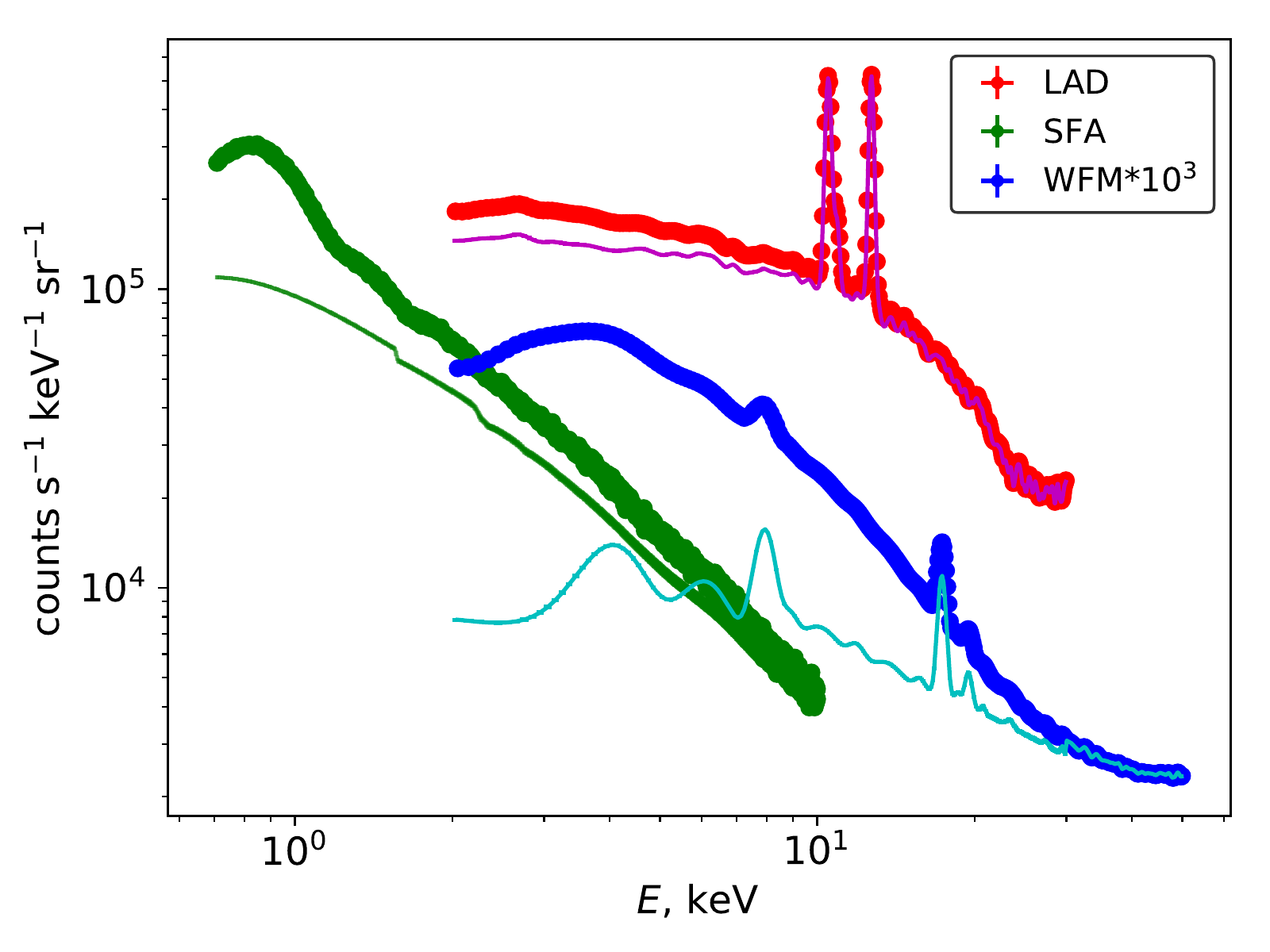}
\includegraphics[width=0.49\linewidth]{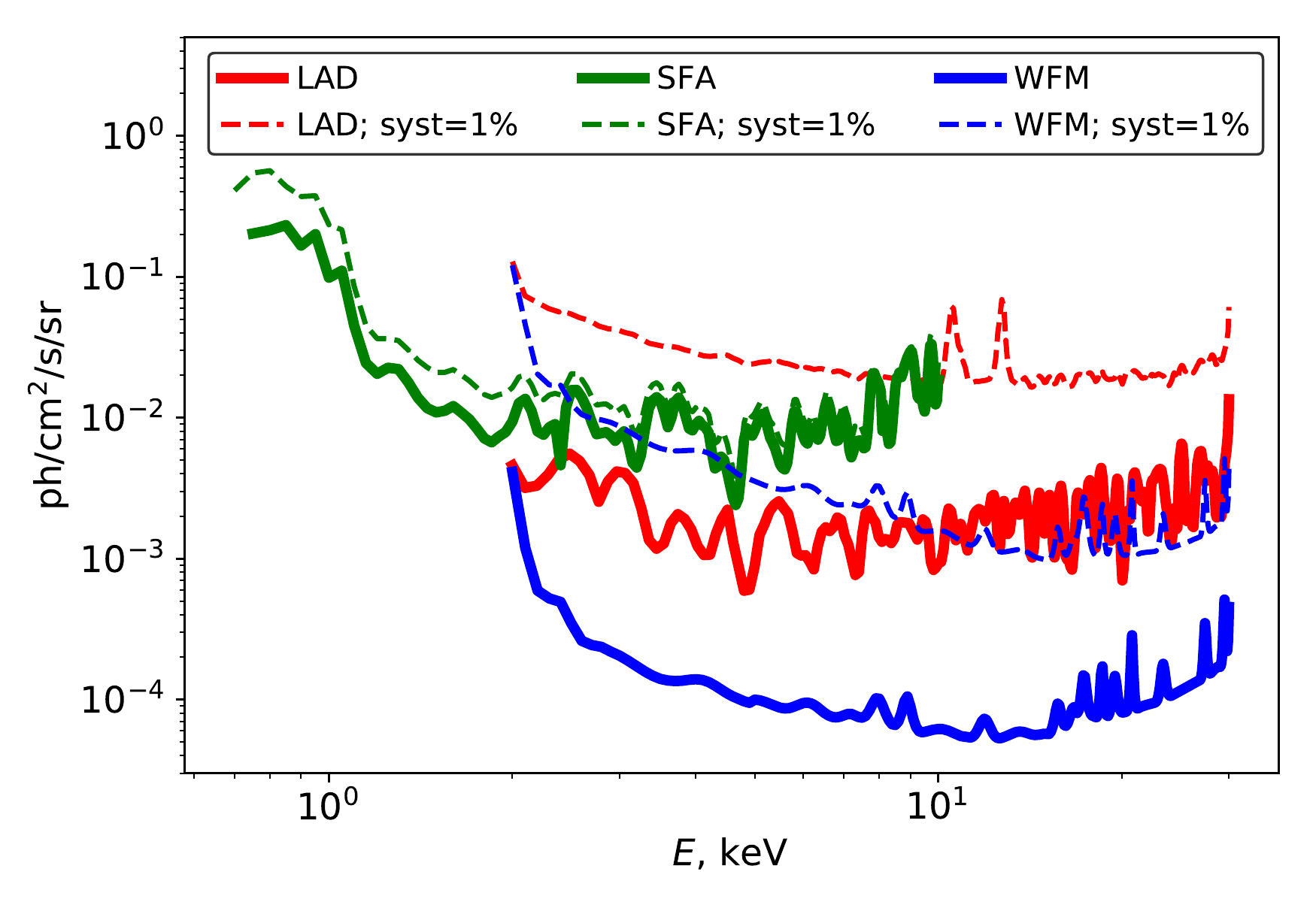}
\caption{\textit{Left:} \extp/LAD, SFA and WFM simulated spectra of 1~Msec observations of a region of blank sky (red and blue points). Cyan, magenta and light green curves illustrate the levels of instrumental background in these instruments. \newline\textit{Right:} Sensitivity of the SFA, LAD and WFM to a narrow Gaussian line present in the whole FoV of the instrument. Dashed lines show the change in the sensitivity of the instrument to the flux, assuming 1\% value of systematic uncertainty.}
\label{fig:spectra_flux_limits}
\end{figure*}

The enhanced X-ray Timing and Polarimetry mission (\extp~\cite{extp_description, extp2, extp1} ) is a forthcoming\footnote{As of 2019 the launch is planned to 2027} Chinese-European mission primarily designed for the study of the equation of state of matter within neutron stars, measurements of QED effects in highly magnetised stars and studies of accretion in the strong-field gravity regime.

The mission will host a set of state of the art scientific instruments operating in the soft to hard X-ray band ($0.5-50$~keV). The main instruments on board the \extp are:\\
-- The Spectroscopic Focusing Array (SFA), consisting of nine X-ray modules operating in the $0.5-10$~keV band with a field of view (FoV) of $12'$ (full width half-maximum, FWHM), total effective area of $\sim 0.8$~m$^2$ at 2~keV and an energy resolution of better than 10\%;\\
-- The Large Area Detector (LAD) -- non-imaging instrument operating at $2-30$~keV energies, with an FoV of $60'$ (FWHM), an effective area of $\sim 3.4$~m$^2$ and an energy resolution better than 250~eV;\\
-- The Wide Field Monitor (WFM) -- a wide, steradian-scale, FoV instrument operating in the $2-50$~keV energy band with an effective area of $\sim 80$~cm$^2$ and an energy resolution similar to that of the LAD. The capabilities of this instrument for indirect dark matter searches were recently discussed by~\cite{zhong20}.

In addition to the instruments described above, the \extp will host another module -- the Polarimetry Focusing Array (PFA). This instrument has a moderate effective area and an energy resolution comparable to current-generation instruments. Thus, in our work, we  will only focus on the prospectives of the SFA, LAD and the WFM for indirect decaying dark matter searches in the keV mass scale. Additional relevant characteristics of these instruments are summarised in Tab.~\ref{tab:instruments}.

\begin{table*}
    \centering
    \begin{tabular}{|c|c|c|c|c|c|}
    \hline
    Instrument & $A_{eff}$, cm$^2$ & $\Omega_{FoV}$, sr & $\Delta E$, keV & $B$,  & $F_{min}^\infty$, \\
    &&&&ph/(cm$^2$s keV sr)&$10^{-2}$ph/(cm$^2$s sr)\\
    \hline
      \extp/LAD   & (23/31/33)$\cdot 10^3$ & $2.4\cdot 10^{-4}$  & 0.29/0.31/0.33 &8.0/5.1/3.4 & 4.6/3.2/2.2\\
      \extp/SFA   & (7.8/5.6/0.9)$\cdot 10^3$  & $9.6\cdot 10^{-6}$ & 0.16/0.16/0.21& 4.5/2.6/5.1 &1.4/0.8/2.1\\
      \extp/WFM   & 42/68/76 & 2.5  & 0.24/0.24/0.26 & 4.8/2.4/0.9 & 2.4/1.2/0.6\\
      \hline\hline
      \xmm/PN & (0.6/0.6/0.1)$\cdot 10^3$\,\,&$4.5\cdot 10^{-5}$\,\,&0.16/0.20/0.28&6.7/5.2/22.2&2.1/2.1/12.4\\
      \textit{Athena}/X-IFU &(6.4/3.5/0.4)$\cdot 10^3$& $3.3\cdot 10^{-6}$\,\, &\,\,(2.6/2.6/3.5)$\cdot 10^{-3}$\,\,&2.8/3.5/26.5 &(1.5/1.8/18)$\cdot 10^{-2}$  \\
      \hline
    \end{tabular}
    \caption{The technical characteristics of the considered \extp instruments compared to the characteristics of the \xmm (PN camera) and  the \textit{Athena}/X-IFU. The table summarises the approximate effective area $A_{eff}$ of each instrument, it's FoV $\Omega_{FoV}$, energy resolution $\Delta E$ (FWHM), the total (instrumental and CXB) expected flux $B$ and the minimal flux detected at infinite exposure assuming 1\% systematic uncertainty ($\alpha=0.01$, see Eq.~\ref{eq:fmin}). Where applicable, the quantities are given at energies 3~keV/5~keV/10~keV and for the \extp's instruments,  derived from the templates/models described in the text. For \xmm and \textit{Athena} missions the quoted parameters were taken from the data used in~\cite{Malyshev_2014, athena_limits}. } 
    \label{tab:instruments}
\end{table*}

\section{Search for decaying DM with \extp}
\label{sec:dm_decay_search}
The flux of a DM-decay line at energy $E=m_{DM}/2$ from an object covering the entire FoV of an instrument is given by
\begin{align}
\label{eq:dm_flux_generic}
& F = \frac{\Gamma}{4\pi m_{DM}}\cdot J_{FoV} \\ \nonumber
& J_{FoV} = \int\limits_{FoV}\int\limits_{l.o.s.}\rho_{DM}d\ell d\Omega 
\end{align}
where $\Gamma$ is the radiative decay width~\cite{pal,barger95} which, for a sterile neutrino, is given by
\begin{align}
\label{eq:gamma}
&  \Gamma = \frac{9\alpha G^2_F}{256 \cdot 4\pi^4}\sin^2(2\theta)m^5_{DM};
\end{align}
$J_{FoV}$ -- is the total J-factor of decaying DM within the field of view; the corresponding integrations are performed over the field of view of the instrument (FoV) and the line of sight distance ($l.o.s.$) to the object. Substituting the expression for $\Gamma$ into Eq.~\ref{eq:dm_flux_generic} one obtains
\begin{align}
\label{eq:dm_flux}
& F_{DM}\approx 10^{-7} \left(\frac{\sin^2(2\theta)}{10^{-11}}\right)\left(\frac{m_{DM}}{10\,\mbox{keV}}\right)^4\times \\ \nonumber
& \times\left(\frac{J_{FoV}}{10^{17}\,\mbox{GeV/cm}^2}\right)\frac{\mbox{ph}}{\mbox{cm}^2\mbox{s}}; 
\end{align}

The J-factor in the direction of a distant object (and consequently its DM-decay signal), is composed of foreground emission from DM present in the Milky Way (MW) galaxy and the signal from DM residing in the source. 

As a matter of fact, within regions of $\sim10'$ (an angular size comparable to FoVs of modern instruments), the DM-decay signal is comparable for a variety of DM-dominated objects with masses ranging from dSphs, to clusters of galaxies ~\cite{dm_in_objects}. Thus, additional considerations such as low levels of astrophysical background, well-measured J-factor, etc., should be taken into account when selecting targets for a deep DM-search observation. On larger scales (specifically $\sim$steradian) the contribution of individual DM-dominated objects becomes negligible in comparison to the expected foreground MW signal. 

Given this, in this study we consider dSphs (for a narrow, $10'$-scale FoV instruments) and MW blank sky regions (for broad, steradian-scale FoV instruments) as the main targets for decaying DM search in the keV band. Contrary to other objects, e.g. clusters of galaxies, in this energy range dSphs and blank sky regions are characterised by low astrophysical backgrounds and can provide a ``clean'' decay-line signal.

The dark matter density profiles for dwarf spheroidal galaxies have been intensively studied in literature~\cite[see e.g.][]{walker09,wolf10,J-factor_values,strigari18}. In our work we rely on numerical J-factors values reported in~\cite{J-factor_values} as a function of the distance from the dSph's center. 

We estimated the MW contribution to the expected signal of a decaying dark matter assuming Navarro-Frenk-White (NFW~\cite{nfw}) profile for dark matter density:
\begin{align}
    \rho_{DM}(r) = \frac{\rho_0 r_0^{3}}{r(r+r_0)^2}
\end{align}
with the $\rho_0=7.8\cdot 10^6\,M_{\bigodot}/\mbox{kpc}^3$, $r_0=17.2$~kpc parameters adapted from the best-fit NFW model of the recent MW-mass distribution study~\cite{mw_dm_profile}; the integration in Eq.~\ref{eq:dm_flux_generic} was performed numerically.

Corresponding values (for both MW and dSph contributions) for the SFA and LAD instruments for a sample of dwarf spheroidal galaxies are summarised in Table~\ref{tab:dsphs}. The uncertainties on J-factors for dSphs illustrate the differences between minimal and maximal J-factor profiles reported in~\cite{J-factor_values}.

Statistics of the DM-decay signal collected within the exposure time $T$ are determined by the line flux (Eq.~\ref{eq:dm_flux}) as well as the intrinsic properties of the instrument. These include, effective area $A_{eff}(E)$, energy resolution $\Delta E$, the level of  background $B$ (instrumental and astrophysical, in ph/(cm$^2$\,s\,keV\,sr) ),  FoV of the instrument $\Omega_{FoV}$ and the level of systematics $\alpha$. The minimal detectable flux of a line scaled to the FoV of the instrument can be estimated as 
\begin{align}
\label{eq:fmin}
& F_{min} = 2\left(\sqrt{\frac{B\Delta E}{A_{eff}T\Omega_{FoV}}}+\alpha B\Delta E\right)\quad \frac{\mbox{ph}}{\mbox{cm}^2\mbox{s sr}}  
\end{align}
where the factor of 2 stands for a $2\sigma$ (or $\sim 95$\% c.l.) detection or upper limit significance. 
Table~\ref{tab:instruments} summarises the basic characteristics of \extp's instruments at 3~keV, 5~keV and 10~keV energies, compared to the characteristics of \xmm and \textit{Athena} missions. The minimal detectable flux in the case of the presence of a 1\% systematic (given by $F_{min}^\infty$ column) allows a rough estimation of the relative sensitivity of the instruments to the narrow-line signal. 

Detailed comparison of $F_{min}(E)$ derived from the data to the expected $F_{DM}(E)$, allows one to derive the range of $(\theta$, $m_{DM})$ values to which the instrument is sensitive.

To perform such a comparison we simulated 1~Msec long observations of a dwarf spheroidal galaxy with both the \extp/SFA and the \extp/LAD instruments. The simulated spectra were assumed to originate from contributions over the whole FoV and to be composed of instrumental and astrophysical background components. 

The instrumental background components were given by the \texttt{XTP\_sfa\_v6.bkg}\footnote{Note, that the provided template corresponds to the background in $\sim 3'$ and has to be re-scaled by a factor of 16 to match $12'$ FoV of SFA. The WFM background template was provided for one module and had to be up-scaled by a factor of 3.}, \texttt{LAD\_40mod\_300eV.bkg} and \texttt{WFM\_M4\_full.bkg} templates for SFA, LAD and WFM respectively, which were provided by the \extp collaboration\footnote{See \href{https://www.isdc.unige.ch/extp/}{\extp website}}. We adopt the FoV size of LAD and SFA instruments from~\cite{extp_description, extp2, extp1}. For the FoV of WFM instrument we adapt a $\Omega_{FoV}=2.5$~sr basing on \texttt{WFM-EXTP\_1OBS\_AREA.fits} spatial template presenting the effective area of a pointing observation. Namely we defined $\Omega_{FoV}\equiv(\sum A_i\Omega_i )/\max(A_i)$, where the sum goes over all pixels of the template and $A_i$, $\Omega_i$ are effective area and the size of $i$-th pixel. The adapted value is within the range ($0.3-4$~sr) quoted in~\cite{extp_description, extp2, extp1} for fully coded and 20\% bounce FoV values.  

For the astrophysical cosmic X-ray background(CXB) for the \extp/LAD and \extp/WFM we adopted a cut-off powerlaw model~\cite{heao_cxb,rxte_cxb,integral_cxb1,integral_cxb2} 
\begin{align}
& F_{CXB} = 7.877E^{-0.29}e^{-E/41.13\,\mbox{keV}}\,\frac{\mbox{keV}}{\mbox{keV\, cm}^2\mbox{s\,sr}}
\end{align}
which well describes the existing data in the $3-60$~keV range.
For the \extp/SFA, which has an energy range extending significantly below 3~keV, we instead adopted the model of CXB derived from \xmm observations of a set of dwarf spheroidal galaxies~\cite{Malyshev_2014}. We verified explicitly that at intersecting energy ranges both models agree within an accuracy of $\sim 10-15$\%.

The observations described above were performed with the \textit{fakeit} XSPEC (version: 12.10.1f) command. The resulting spectra (normalised per FoV of corresponding instrument) are shown in the left panel of Fig.~\ref{fig:spectra_flux_limits}. Red, green and blue points illustrate the total expected flux seen by LAD, SFA and WFM correspondingly, while magenta, light-green and cyan lines present the level of the instrumental background.

We would like to note that the instrumental background of \extp strongly varies between the instruments. The SFA's background is featureless and can be adequately modelled by a sum of two powerlaw models (convolved and not convolved with the effective area). The background of WFM below 30~keV can be modelled with a broken powerlaw model, containing a break at $E_{br}\sim 16$~keV, and hosts multiple instrumental lines. To avoid further complications with the background model of this instrument hereafter, we limit the considered energy range for this instrument to $2-30$~keV. Finally, the instrumental background of the LAD is even more complicated and can not be modelled accurately with any simple model.

\subsection{Observational strategy}
Given these points, we propose somewhat different observational strategies of a dSph by the SFA and LAD instruments. For the SFA we propose that the observation should be centered on the dSph and accompanied with subsequent modelling of instrumental and astrophysical background. Thus, a DM-decay line can be searched for on top of the modelled background. This strategy is similar to one widely used in decaying dark matter searches in astrophysical objects, see e.g.~\cite{sterile_neutrino_review19} for a review. 

For the LAD, we propose performing a set of ``ON-OFF'' observations, where ``ON''-observations are centered on the dSph and ``OFF'' -- on an empty sky region close to the object, but for which the contribution from dSph DM-decay signal is minimal. In this case we propose that rather than modelling astrophysical/instrumental backgrounds, to instead use ``OFF'' observations as a background for ``ON'' observations. 
The DM-decay line in this case is searched for in the obtained, background subtracted, (consistent with 0) spectrum. Such a strategy allows one to avoid modelling the complex LAD background and/or potential systematic effects connected with our poor knowledge of it.

The extremely large FoV ($\sim 2.5$~sr) of the WFM instrument unavoidably covers a region much broader than the angular size any known dSph, and therefore the contribution to the expected signal of any dSph in this FoV will be negligible. To fully utilise the capacity of the WFM in dark matter searches, we propose instead use it to observe blank sky regions characterised by low astrophysical background.

We note that in case of blank sky observations with WFM the ``ON-OFF'' strategy is only marginally possible since the expected dark matter signal is by an order of magnitude comparable in any direction on the sky similarly to possible variations of the astrophysical background. Yet, to maximise the expected signal within ``ON-OFF'' strategy one may locate the ``ON'' region as close as possible to the Galactic Center (as was proposed e.g. by~\cite{zhong20}). We note, however that in this case an additional astrophysical component -- galactic bulge/ridge X-ray emission (GRXE) should be taken into account. The GRXE emission is present at low galactic latitudes and is believed to originate from a population of unresolved X-ray binaries~\cite{krivonos07}. At these latitudes, GRXE flux can exceed the flux of cosmic X-ray background by an order of magnitude~\cite{krivonos07,perez19}. It's spectrum is not featureless and hosts multiple astrophysical lines at least at energies $\lesssim 3$~keV~\cite{ebisawa08} which can lead to additional confusion between astrophysical and DM-decay signal. 

To minimise potential GRXE contribution we propose to observe relatively high galactic latitudes ($|b|>20$) with the WFM, where the GRXE contribution is minimal~\cite{krivonos07}. We propose also to locate the quasi-rectangular FoV of the WFM parallel to the galactic plane to minimise the average distance to the Galactic Center and thus maximise the expected DM-decay signal.

\subsection{Results}
Following the proposed strategy for the SFA and WFM, we perform a search for a narrow Gaussian line originating from the whole FoV, on top of the modelled backgrounds (specifically, the sum of the instrumental and astrophysical background models as described above). For the LAD we performed an additional 1~Msec long simulation of an ``OFF'' region characterised by the same astrophysical/instrumental backgrounds as an ``ON'' observation of a dSph. In this case we performed the search for a narrow Gaussian line in the background subtracted spectrum. Upper limits of $2\sigma$ ($\sim 95\%$ confidence level) on the normalisation of such line\footnote{The upper limits were calculated with the \texttt{error 4.0 } XSPEC command.} are shown with solid blue (SFA) and red (LAD) curves in the right panel of Fig.~\ref{fig:spectra_flux_limits}. These limits are exact equivalents of the minimal detectable flux in Eq.~\ref{eq:fmin}.

We would like to stress the significance of the potential effects of systematic uncertainties on the limits which can be derived by the LAD and WFM instruments. To simulate this effect we modified \texttt{STAT\_ERR} column of simulated spectral files by adding a value proportional to the total number of counts observed each channel. Dashed curves in the right panel of Fig.~\ref{fig:spectra_flux_limits} present limits on the line normalisation which can be obtained in the presence of a 1\%  systematic uncertainty. We conclude that in the case where \extp systematic is not well controlled, the subsequent limits for decaying DM by the LAD and WFM will worsen by a factor of $\gtrsim 10$. On the contrary, the low instrumental background and relatively small FoV of the SFA do not allow statistical uncertainty to substantially overcome systematic uncertainty in  within a 1~Msec observation. Consequently presented limits are only weakly dependent on any added systematics.

\begin{table}[t!]
\begin{tabular}{|c | c | c | c |}
\hline
dSph                 & Galactic  &  $J_{FoV}(6')$ &   $J_{FoV}(30')$                \\
                     & coordinates&  $10^{17}$~GeV/cm$^2$ &   $10^{17}$~GeV/cm$^2$                \\
\hline
Segue 1 & (220.5; 50.4) & $0.84 + 2.0^{+2.1}_{-1.2}$  & $9.8^{+14.8}_{-8.4}$  \\
Draco   & (86.4; 34.7)  & $1.1  + 2.2^{+0.6}_{-0.5}$      & $33.4^{+17.8}_{-16.0}$ \\
Carina  & (260.1; -22.2) & $1.0 + 0.9^{+0.2}_{-0.1}      $      & $  7.9^{+7.4}_{-4.1} $ \\
Fornax   & (237.1; -65.7)  & $0.95 + 1.0^{+0.3}_{-0.2}$      & $ 7.2^{+1.7}_{-1.4}  $ \\
Sextans   & (243.5; 42.3 )  & $0.90 + 0.5^{+1.0}_{-0.2}$      & $ 7.8^{+6.3}_{-5.3}  $ \\
Sculptor  & (287.5; -83.2)  & $1.1 + 1.7^{+0.3}_{-0.3} $      & $ 15.5^{+5.9}_{-3.9} $ \\
Ursa Minor& (105.0; 44.8 )  & $1.0 + 2.4^{+0.9}_{-0.8} $      & $ 10.7^{+14.2}_{-4.4}$ \\
Ursa Major I& (159.4; 54.4) & $0.8 + 0.7^{+1.0}_{-0.4} $      & $ 4.1^{+7.3}_{-3.2}  $ \\
Ursa Major II& (152.5; 37.4)& $ 0.75 + 2.3^{+3.7}_{-1.5}$     & $ 23.9^{+48.3}_{-18.1}$\\
Bootes I  & (358.0; 69.6 )  & $1.4 + 0.9^{+0.9}_{-0.5}$      & $    8.0^{+13.7}_{-6.3}$ \\
Coma Ber  & (241.9; 83.6 )  & $1.1 + 1.8^{+2.1}_{-1.0}$      & $ 9.2^{+13.9}_{-6.8} $ \\
\hline
\end{tabular}
\caption{Parameters of a sample of dSph galaxies. J-factor in the field of view of SFA ($J_{FoV}(6')$) is given as a sum of Milky Way~\cite{mw_dm_profile} and dSph~\cite{J-factor_values} contributions. J-factors for the field of view of the LAD ($J_{FoV}(30')$) correspond only to contributions from dSphs, see text for further details. Uncertainties on dSph contributions illustrate the values for minimal and maximal expected J-factor within the selected radius. }
\label{tab:dsphs}
\end{table}

Using the derived results for the \extp's sensitivity to a narrow Gaussian line, we obtain the corresponding minimal value of the mixing angle $\theta$ at which a DM-decay line can be detected at a given energy $E=m_{DM}/2$. Corresponding limits for 1~Msec long Segue~1 dSph observations (assumed $J_{FoV}=2.84\cdot 10^{17}$~GeV/cm$^2$ for the SFA ; $J_{FoV}=9.8\cdot 10^{17}$~GeV/cm$^2$ for the LAD and $J_{FoV}=2\cdot 10^{22}$~GeV/cm$^2$ for the WFM observations of a blank sky region centered at Segue~1 and parallel to the galactic plane ) are shown in Fig.~\ref{fig:dm_limits} along with current theoretical and observational constraints of sterile neutrino parameters (see e.g.~\cite{sterile_neutrino_review19} for the review). Also displayed for comparison are the expected limits on observations by the forthcoming \textit{Athena} mission~\cite{athena_limits},given the same exposure and target. Note that here, the presented limits correspond to a zero level of systematic uncertainty. The expected limits from observations of other low astrophysical background DM-dominated objects, can be obtained by re-scaling presented limits according to the $J_{FoV}$ of the target, see e.g. Table~\ref{tab:dsphs}.

\section{Discussion and Conclusions}
\begin{figure*}[t!]
\includegraphics[width=0.49\linewidth]{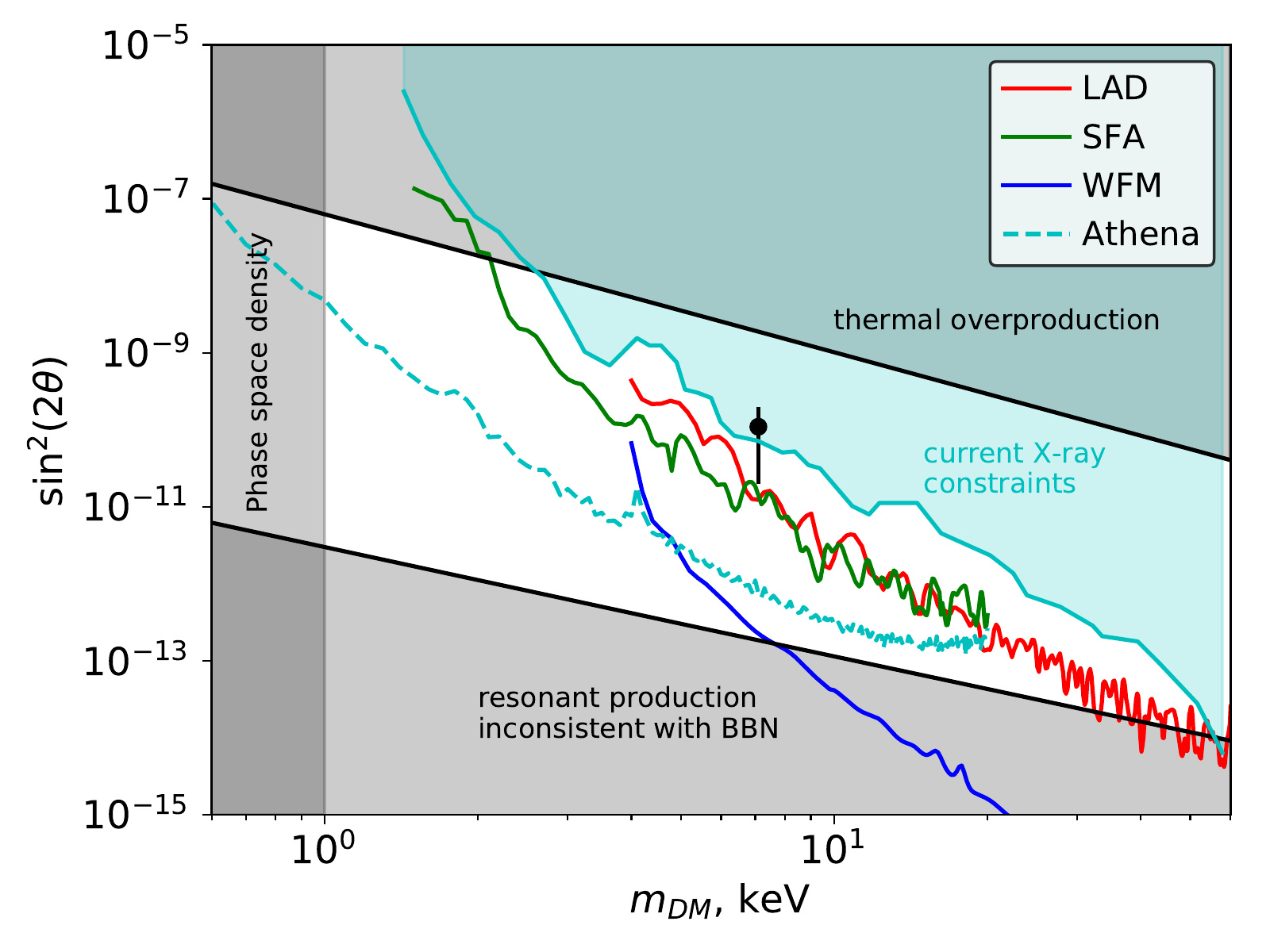}
\includegraphics[width=0.49\linewidth]{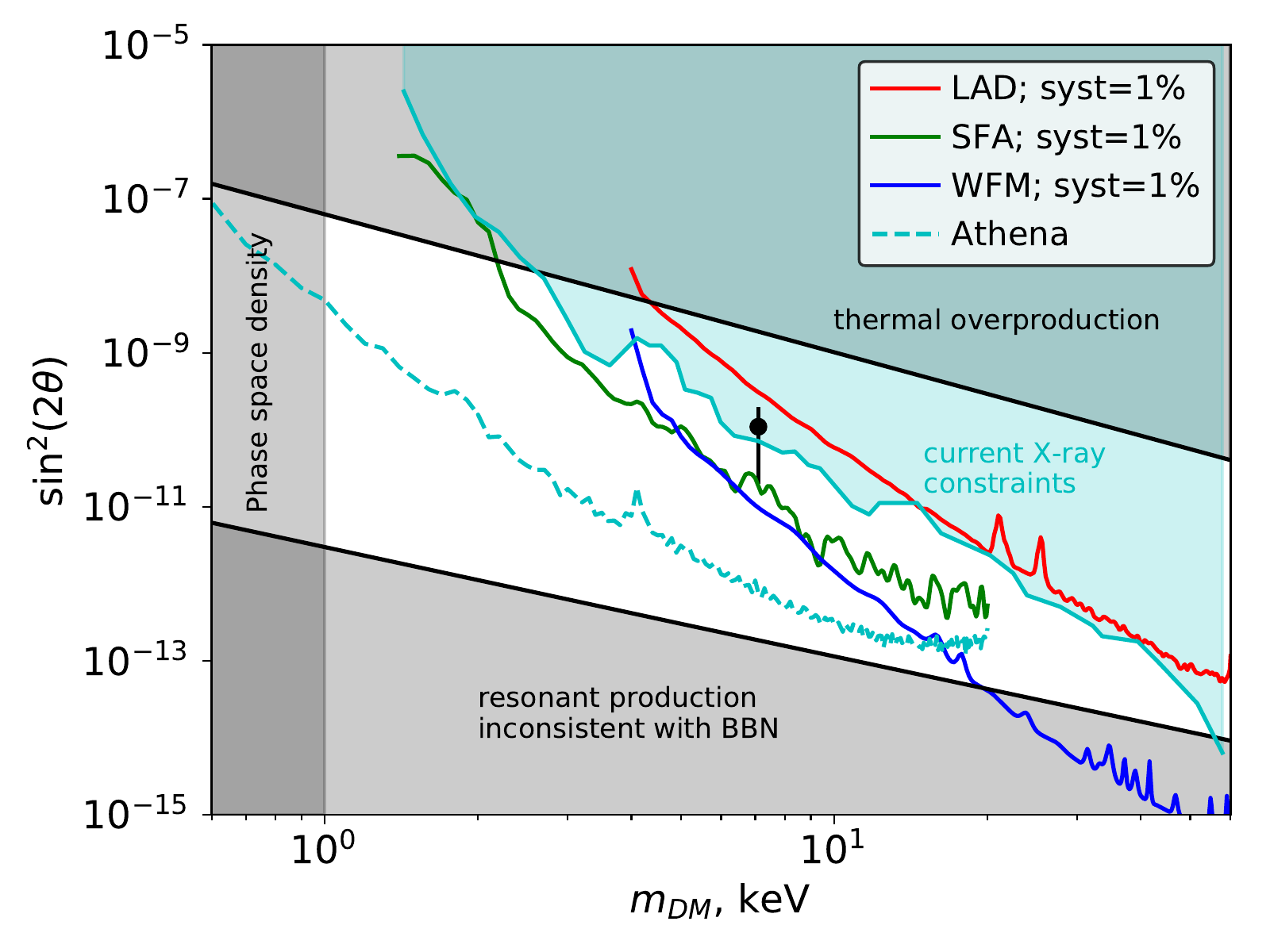}
\caption{$2\sigma$ sensitivity reach of the \extp to the parameters of the sterile neutrino from 1~Msec observations of Seg I dSph (by LAD and SFA) and same duration blank sky observations (WFM). Left and right panels assume a zero and 1\% level of systematic uncertainty for all instruments correspondingly. Used J-factors correspond to mean values reported in Tab.~\ref{tab:dsphs} and in the text. The cyan dashed curve illustrates $2\sigma$ Athena constraints from 1~Msec observations of the same target~\cite{athena_limits}. The light blue region shows the existing constraints (adapted from~\cite{sterile_neutrino_review19}). Phase space density~\cite{tremainegunn,tg1,gorbunov08,savchenko19}, thermal overproduction~(see \cite{dodelson} and \cite{bertone05,sterile_neutrino_review19} for the review ) and the bounds originating from the abundances of light elements produced during BBN~\cite{serpico05} are shown as grey regions. The black point represents the sterile neutrino parameters from the tentative detection of an unidentified $\sim 3.55$~keV line in certain DM-dominated objects~(see \cite{line35},\cite{line35_1} and \cite{sterile_neutrino_review19} for a recent review) }
\label{fig:dm_limits}
\end{figure*}

This study has demonstrated the capability of the upcoming \extp satellite in searching for decaying dark matter and found it can impose significantly better limits than current observational means. Observations with \extp of DM-dominated objects with exposures of 1Msec e.g. Segue~1 (by SFA and LAD), or blank sky regions (by WFM), have the potential to improve existing $2\sigma$ X-ray observational constraints by a factor of $\sim 5-10$ within the $2-50$~keV dark matter particle mass range (see Fig.~\ref{fig:dm_limits}, right panel), assuming 1\% level of systematic uncertainty. The same constraints for the un-likely case of much smaller, consistent with zero, level of systematic are significantly better and are shown in the left panel of Fig.~\ref{fig:dm_limits}.

We assert that the systematic uncertainty will play a significant role in constraining decaying DM parameters from LAD and WFM data. Uncontrolled systematics at a level of $\gtrsim 1$\% can detrimentally affect obtained constraints by an order of magnitude in comparison to zero-systematic case. In the case of the LAD  this could produce constraints comparable to, or even worse than, those of current X-ray instruments. When considering the proposed 1~Msec observation, the low instrumental background and relatively narrow FoV of the SFA makes the effects of systematics less significant in this instrument. The systematic at a level of 1\% (comparable to the estimated flux systematic uncertainty of \xmm\footnote{See e.g. \href{http://xmm2.esac.esa.int/docs/documents/CAL-TN-0018.pdf}{EPIC Calibration Status document}}) will lead to a deterioration of zero-systematic constraints by only a factor of $\sim 1.5$.

We note that, a 1\% systematic uncertainty can be a reasonable estimation for \xmm-like instruments such as SFA and LAD. However, for a broad-FoV instrument not designed specifically for spectral studies such as WFM, we recognise that this uncertainty could be rather optimistic.

The constraints presented in Fig.~\ref{fig:dm_limits} indicate also that the \extp will be sensitive enough to exclude or detect, at $3\sigma$ level, a sterile neutrino with the mass of $m_{DM}\sim 7$~keV and a mixing angle of ($\sin^2(2\theta)\sim 2\cdot 10^{-11}$). This angle roughly corresponds to the minimal mixing angle of a sterile neutrino producing a $\sim 3.55$~keV line, as discussed in literature. This line has been tentatively detected in some DM-dominated objects and is still actively being discussed in the field (see~\cite{line35_1, line35} and \cite{sterile_neutrino_review19} for a recent review). The corresponding range of mixing angles discussed is denoted by the black point with error-bars in Fig.~\ref{fig:dm_limits}. 

With the optimistic assumptions on the mixing angle $\sin^2(2\theta)\sim 8\cdot 10^{-11}$ (corresponding to $2\sigma$ limits on mixing angle from current X-ray observations), the DM-decay line can be detected with a significance of $\gtrsim 10\sigma$, given a 1\% systematic with \extp/SFA or WFM (in line with estimations of~\cite{zhong20}) instruments. The strength of such a significant line could be compared across the sample of other DM-dominated objects and/or along the sky in order to correlate its intensity with the known $J_{FoV}$ value and thus draw conclusions on its DM-decay origin.

Alongside its numerous other scientific objectives, \extp will be a precursor to the forthcoming \textit{Athena} mission's decaying dark matter searches. The improved sensitivity of \extp in comparison to the current generation of instruments will lead to a significant reduction of the sterile neutrinos unobserved parameter space. We assert that with well controlled systematic uncertainties, the \extp has the potential to discover decaying dark matter and make the first estimations of its parameters which can be further verified with Athena.

\noindent\textit{Acknowledgements}
The authors acknowledge support by the state of Baden-W\"urttemberg through bwHPC. This work was supported by DFG through the grant MA 7807/2-1.

\def\aj{AJ}%
\def\actaa{Acta Astron.}%
\def\araa{ARA\&A}%
\def\apj{ApJ}%
\def\apjl{ApJ}%
\def\apjs{ApJS}%
\def\ao{Appl.~Opt.}%
\def\apss{Ap\&SS}%
\def\aap{A\&A}%
\def\aapr{A\&A~Rev.}%
\def\aaps{A\&AS}%
\def\azh{AZh}%
\def\baas{BAAS}%
\def\bac{Bull. astr. Inst. Czechosl.}%
\def\caa{Chinese Astron. Astrophys.}%
\def\cjaa{Chinese J. Astron. Astrophys.}%
\def\icarus{Icarus}%
\def\jcap{J. Cosmology Astropart. Phys.}%
\def\jrasc{JRASC}%
\def\mnras{MNRAS}%
\def\memras{MmRAS}%
\def\na{New A}%
\def\nar{New A Rev.}%
\def\pasa{PASA}%
\def\pra{Phys.~Rev.~A}%
\def\prb{Phys.~Rev.~B}%
\def\prc{Phys.~Rev.~C}%
\def\prd{Phys.~Rev.~D}%
\def\pre{Phys.~Rev.~E}%
\def\prl{Phys.~Rev.~Lett.}%
\def\pasp{PASP}%
\def\pasj{PASJ}%
\def\qjras{QJRAS}%
\def\rmxaa{Rev. Mexicana Astron. Astrofis.}%
\def\skytel{S\&T}%
\def\solphys{Sol.~Phys.}%
\def\sovast{Soviet~Ast.}%
\def\ssr{Space~Sci.~Rev.}%
\def\zap{ZAp}%
\def\nat{Nature}%
\def\iaucirc{IAU~Circ.}%
\def\aplett{Astrophys.~Lett.}%
\def\apspr{Astrophys.~Space~Phys.~Res.}%
\def\bain{Bull.~Astron.~Inst.~Netherlands}%
\def\fcp{Fund.~Cosmic~Phys.}%
\def\gca{Geochim.~Cosmochim.~Acta}%
\def\grl{Geophys.~Res.~Lett.}%
\def\jcp{J.~Chem.~Phys.}%
\def\jgr{J.~Geophys.~Res.}%
\def\jqsrt{J.~Quant.~Spec.~Radiat.~Transf.}%
\def\memsai{Mem.~Soc.~Astron.~Italiana}%
\def\nphysa{Nucl.~Phys.~A}%
\def\physrep{Phys.~Rep.}%
\def\physscr{Phys.~Scr}%
\def\planss{Planet.~Space~Sci.}%
\def\procspie{Proc.~SPIE}%
\let\astap=\aap
\let\apjlett=\apjl
\let\apjsupp=\apjs
\let\applopt=\ao
\bibliography{biblio}

\end{document}